\documentclass{article}
\usepackage{amssymb}
\usepackage{cite,multirow} 
\usepackage[dvips]{graphicx,epsfig}
\usepackage{graphics}
\def\1ad{\mbox{\normalsize $^1$}}
\def\2ad{\mbox{\normalsize $^2$}}
\def\3ad{\mbox{\normalsize $^3$}}
\def\4ad{\mbox{\normalsize $^4$}}
\def\5ad{\mbox{\normalsize $^5$}}
\def\6ad{\mbox{\normalsize $^6$}}
\def\7ad{\mbox{\normalsize $^7$}}
\def\8ad{\mbox{\normalsize $^8$}}

\makeatletter
\@addtoreset{equation}{section}
\makeatother

\begin{document}

\rightline{hep-th/0512240}
\rightline{Bicocca-FT-05-27}
\vskip 0.3cm
\centerline{{ \bf \Large From Toric Geometry to Quiver Gauge Theory: }}
\vskip 0.1cm
\centerline{{ \bf \Large the Equivalence of a-maximization and Z-minimization}}
\vskip 0.4cm
\centerline{{\bf Agostino Butti 
 and Alberto Zaffaroni}}
\vskip 0.4cm
\begin{center}
\em {Dipartimento di Fisica, Universit\`{a} di Milano-Bicocca, \\ 
P.zza della Scienza, 3; I-20126 Milano, Italy.}\\[0.3em]
E-mail:~\textsf{agostino.butti@mib.infn.it} $\quad$ \textsf{alberto.zaffaroni@mib.infn.it}
\vskip .4cm

\end{center}




\vskip 1.0cm

\begin{abstract}
\noindent
AdS/CFT predicts a precise relation between the central charge $a$, the scaling dimensions of some operators in the CFT on D3-branes at conical singularities and the volumes of the horizon and of certain cycles in the supergravity dual. We review how a quantitative check of this relation can be performed for all toric singularities. In addition to the results presented in hep-th/0506232, we 
also discuss the relation with the  recently discovered map between toric singularities and tilings; in particular, we discuss how to find the precise distribution of R-charges in the quiver gauge theory using dimers technology.

\vskip 0.5em

\noindent
{\em To appear in the proceedings of the RTN workshop: ``Constituents, Fundamental Forces and Symmetries of the Universe'', Corfu, Greece, 20-26 Sept. 2005.}
\end{abstract}


\large
\section{Introduction}

D3 branes living at the singularity of a Calabi-Yau cone have provided general and interesting results 
for the AdS/CFT correspondence since its early days.
The IR limit of the gauge theory on the world volume of the D3-branes
is dual to type IIB string theory on the near horizon geometry 
$AdS_5\times H$, where the horizon manifold $H$ is the compact base of the cone \cite{kw,horizon}. Since the cone is Calabi-Yau, $H$ has a Sasaki-Einstein metric.    
Until few months ago, the only known Sasaki-Einstein metrics
were the round sphere $S^5$ and $T^{1,1}$, the horizon of the conifold.
Recently, various infinite classes of new Sasaki-Einstein metrics 
were constructed \cite{gauntlett,CLPP,MSL} and named $Y^{\bar p,\bar q}$
and $L^{p,q,r}$ and the corresponding dual gauge theories were determined 
\cite{benvenuti,kru2,noi,tomorrow}. 

The remarkable growth in the number of explicit examples was accompanied by a deeper general understanding of the correspondence, in particular when the Calabi-Yau cone is a toric manifold. The AdS/CFT correspondence predicts a precise relation between the central charge $a$, the scaling dimensions of some operators in the CFT on the D3-branes  and the volumes of $H$ and of certain submanifolds. 
Checks of this relation have been done for the known Sasaki-Einstein metrics
\cite{benvenuti,bertolini,kru,kru2,noi,tomorrow}.
It is by now clear that all these checks can be done without an
explicit knowledge of the metric. a-maximization \cite{intriligator} provides
an efficient tool for computing central and R-charges on the quantum field theory side. On the other hand, Z-minimization \cite{MSY} 
provides a geometrical method for extracting volumes from the toric data. In 
this note we review the proof of the equivalence between a-maximization and
Z-minimization given in \cite{aZequiv} \footnote{For other interesting and complementary results on the equivalence based on supergravity see \cite{Tachikawa,Barnes}.}. In this process, we provide a general formula for assigning R-charges and multiplicities for the chiral fields of the quiver gauge theory based only on toric data. Our result is a quantitative check of the AdS/CFT correspondence that can be performed for all toric manifolds.

Since our work \cite{aZequiv} appeared, the  long standing problem of  finding the correspondence between toric singularities and quiver gauge theories has been completely solved
using dimer technology. The brane tilings \cite{dimers} provide an ingenious way of reconstructing the toric cone from the quiver; the inverse, and more complicated, problem of computing the gauge theory from the toric 
diagram has been recently solved using zig-zag paths in \cite{rhombi}, and nicely interpreted in \cite{mirror}. 
Now that we have an algorithm for determining the quiver gauge theory from the toric data, it is important
to study the distribution of R-charges among the chiral fields of the gauge theory. This information is not
important for determining the value of the central charge and the R-charges but it is certainly important
for a better understanding of the CFT. In this note we discuss in details how to find the precise distribution of R-charges in the quiver gauge theory using dimers and zig-zag paths.

\section{The gauge theory and AdS/CFT predictions}
\label{gauge}
We consider $N$ D3-branes living at the tip of a CY cone.
The base of the cone, or horizon, is a five-dimensional Sasaki-Einstein manifold $H$ 
\cite{kw,horizon}. The ${\cal N}=1$ gauge theory living on the branes is 
superconformal and dual in the AdS/CFT correspondence to the type IIB
background $AdS_5\times H$, which is the near horizon geometry.

As well known, the matter content of the gauge theory can be represented
by a quiver diagram, where each node represents a gauge group and oriented 
links represent chiral bifundamental multiplets. To complete the description 
of the gauge theory one must specify also the superpotential.
By applying Seiberg dualities to a quiver gauge theory we can obtain
different quivers that flow in the IR to the same CFT.
It turns out that one can always find phases where all the gauge groups have the same number of colors; these are called toric phases.
For toric phases, and when the dual geometry is toric, one can ``lift'' the quiver diagram and 
draw it on a torus $ T^2$ \cite{dimers}. This diagram, called the periodic quiver, identifies completely
the gauge theory, since now every superpotential term in the gauge theory is described by a face: it is the trace of the product of chiral
fields of the face \footnote{The superpotential has a sign + or - if the arrows of the face in the periodic quiver are oriented
clockwise or anticlockwise respectively. In the dual graph, the dimer,  a white or black vertex correspond to a term with sign + or - respectively.}. The dual graph of the periodic quiver, known as the brane tiling or dimer configuration,
is still drawn on a torus $T^2$ (look at Figure \ref{y21} for the example $Y^{2,1}$). In the dimer the role of faces and vertices is exchanged: faces are gauge groups and vertices are superpotential terms. The dimer is a bipartite graph: it has an equal number of white and black vertices and links connect only vertices of different colors.

The first prediction of the correspondence we want to check is the relation between 
the central charge in field theory
and the volume of the internal manifold \cite{gubser}:
\begin{equation}
a=\frac{\pi^3}{4 {\rm Vol}(H)}
\label{central}
\end{equation}
The second quantitative prediction states that the exact R-charges of chiral fields $\phi_i$
are proportional to volumes of certain calibrated 3d submanifolds $\Sigma_i$ inside $H$ \cite{gubserkleb}:
\begin{equation}
R_i=\frac{\pi {\rm Vol}(\Sigma_i)}{3 {\rm Vol}(H)}
\label{baryons}
\end{equation}
Recall in fact that the baryon operator built with $\phi_i$ is dual to a D3 brane wrapped
over $\Sigma_i$ and equation (\ref{baryons}) is the relation between the scaling dimension of the operator and
the mass of the state in the string dual.

\section{a-maximization}
\label{amax}

We will denote with $V$ the number of vertices of the periodic quiver (gauge groups), 
with $E$ the number of edges (chiral fields), and with $F$ the number of faces (superpotential terms).
Since the periodic quiver is drawn on a torus, the Euler relation implies \cite{dimers}:
\begin{equation}
V+F=E
\label{eul}
\end{equation}

The R-charges $R_i$ of the chiral fields $\phi_i$, $i=1,\ldots E$, satisfy a set of linear constraints: 
since we are looking for the infrared fixed points of the gauge theory,
each term in the superpotential must have R-charge $2$ ($F$ equations) and the 
exact NSVZ $\beta$ function for every gauge group must be zero ($V$ equations).
The latter condition coincides with the anomaly cancellation for the R-symmetry.

It seems from (\ref{eul}) that $F+V$ linear conditions will determine uniquely the $E$ unknown charges
$R_i$. However, in the cases we are interested in, the conditions 
are not linearly independent: it is easy to see that the homogeneous part of the linear system is just solved by the global 
non anomalous $U(1)$ charges for chiral fields, so that a generic R-symmetry is mixed with the global non anomalous 
symmetries:
\begin{equation}
R_i=R_i^0+x_h S_i^h,\qquad \qquad h=1,\ldots d-1
\label{poss}
\end{equation} 
where $R^0$ is a particular solution, the $S^h$ are the global symmetries, and $x_h$ are real parameters.

The important fact is that, in general, we expect $d-1$ global non anomalous symmetries, where $d$ is the number of
sides of the toric diagram in the dual theory.
We can count these symmetries from the 
number of massless vectors in the $AdS$ dual. Since the manifold is toric, 
the metric has three $U(1)$ isometries.
One of these (generated by the Reeb vector) corresponds to the R-symmetry while the other 
two give two global flavor symmetries in the gauge theory. 
Other gauge fields in $AdS$ come 
from the reduction of the RR four form on the non-trivial three-cycles
in the horizon manifold $H$, and there are $d-3$ three-cycles in homology \cite{tomorrow}  when $H$ is smooth.
On the field theory side,
these gauge fields  correspond to baryonic symmetries.
Summarizing, the global non anomalous symmetries are:
\begin{equation}
U(1)^{d-1}=U(1)^2_F \times U(1)^{d-3}_B
\end{equation}

At the fixed point, 
only one of the possible non-anomalous R-symmetries (\ref{poss}) enters
in the superconformal algebra. It is the one in the same multiplet as the 
stress-energy tensor.
The value of the exact R-charges at the fixed point can 
be found by using the a-maximization technique \cite{intriligator}. As shown in
\cite{intriligator}, we have to maximize the trial a-charge \cite{anselmi}:
\begin{equation}
a(R)=\frac{3}{32}(3 {\rm Tr} R^3-{\rm Tr} R)
\label{ans}
\end{equation}
when R ranges over the possible symmetries (\ref{poss}); the value of this function at the
maximum gives the central charge, and the position of the maximum gives the exact R-charges.
In (\ref{ans}) the trace is a sum over all the fermionic components of the multiplets.

\section{Multiplicities and R-charges from the toric diagram}
\label{geometry}
 
\begin{figure}
\begin{minipage}[t]{0.48\linewidth}
\centering
\includegraphics[scale=0.60]{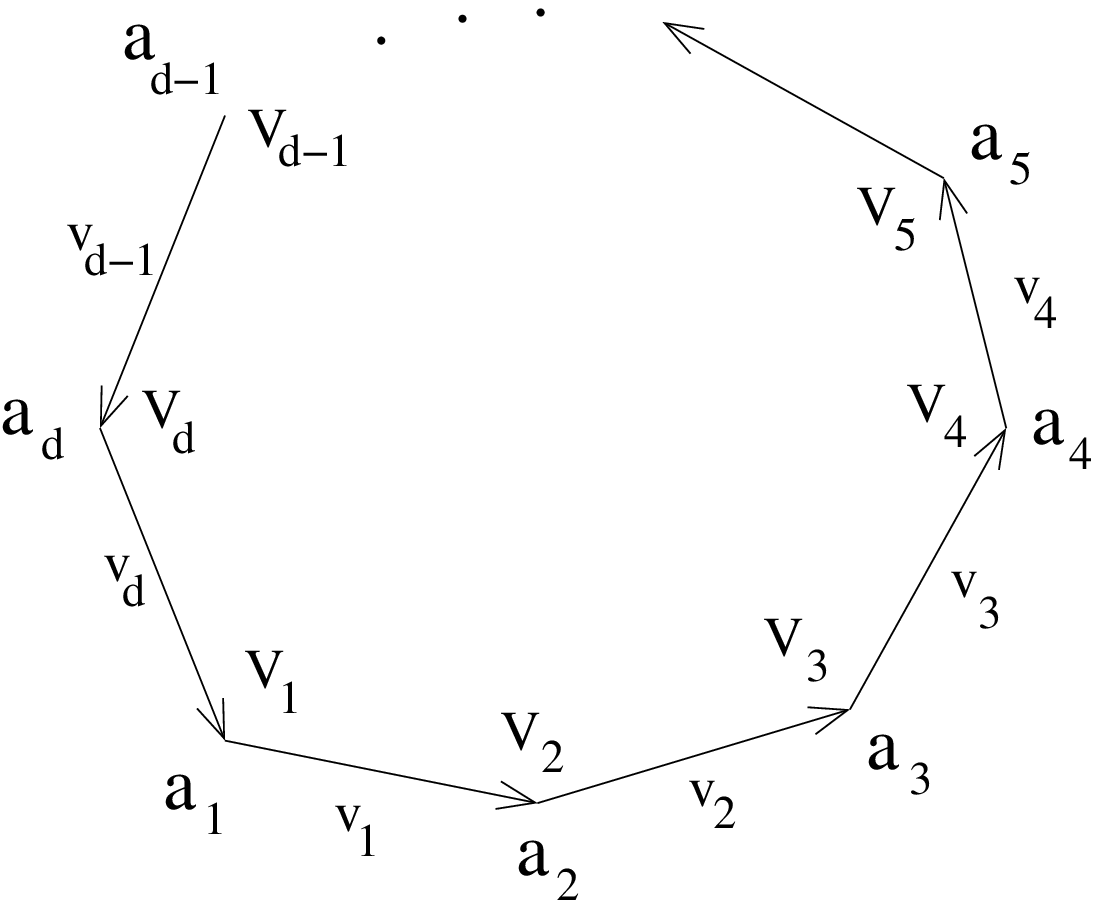}
\caption{The convex polygon $P$.}
\label{polygon}
\end{minipage}%
~~~~~~\begin{minipage}[t]{0.48\linewidth}
\centering
\includegraphics[scale=0.60]{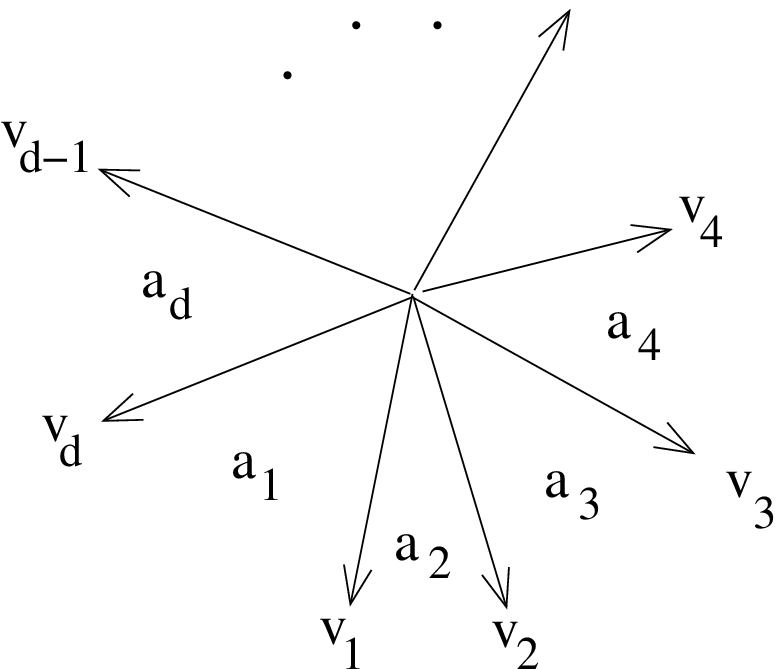}
\caption{The $(p,q)$ web for $P$.}
\label{pqweb}
\end{minipage}%
\end{figure}

Every toric CY cone in six dimensions is described by a toric diagram $P$ \footnote{The toric diagram is the intersection of the fan with the plane where all the generators live, due to the Calabi-Yau condition. 
For the necessary elements of toric geometry see \cite{fulton} and the review part of \cite{MS}.},
which is simply a convex polygon in the plane with integers vertices (Figure \ref{polygon}), encoding in a simple diagrammatic way many information about the toric action. Toric diagrams equivalent up to translations and $SL(2,\mathbb{Z})$ transformations describe the same manifold. Moreover for every convex polygon with integer vertices there exists a unique singular Calabi-Yau metric over the corresponding toric cone. Therefore toric diagrams identify completely the geometry.
If the sides of the toric diagram do not pass through integer points the horizon $H$ is smooth; here and in the following we shall restrict to this case (but it is not difficult to extend the results to the general case \cite{aZequiv}).

Another fact from toric geometry we shall use is that there is a correspondence between vertices $V_i$ of the toric diagram $P$ and calibrated 3d submanifolds $\Sigma_i$ of $H$ (those appearing in equation (\ref{baryons})).

It is sometime useful to think in terms of $(p,q)$ webs.
The $(p,q)$ web is obtained by taking the perpendiculars to the vectors $v_i$ of the toric diagram with the same length as in $P$, see Figure \ref{pqweb} \footnote{With a little abuse of notation we call $v_i$ both the sides of $P$ and the vectors of the $(p,q)$ web. In fact they differ only by a rotation of $90^o$.}. 
Let us also define the symbol:
\begin{equation}
\langle w_i, w_j \rangle \equiv \det(w_i, w_j) 
\end{equation} 
where $w_i$ and $w_j$ are two vectors in the plane of $P$.

Some of the data of the gauge theory can be extracted directly from the
geometry of the cone. In particular, there exist simple formulae for
the number of gauge groups $V$ and the total 
number of chiral bi-fundamental fields $E$ \cite{hananymirror,benvenuti}
\begin{eqnarray}
V&=& 2{\rm Area}(P)\nonumber\\
E&=& \frac{1}{2}\sum_{i,j} |\langle v_i,v_j\rangle |
\label{numbers}
\end{eqnarray}
We would like to stress that the expression for $E$ is not true in all toric phases, but remarkably, 
for the gauge theories in consideration, it seems that it exists always one or more  toric phases where this formula is true. 
We will call these phases minimal, since they are the ones with a minimal
 number of fields.
We shall consider here only minimal toric phases (for a conjectured extension to non minimal toric phases see \cite{aZequiv}).
 
Generalizing results in the literature similar to (\ref{numbers}), in particular see \cite{hananymirror,kru2,benvenuti,tomorrow}, the following algorithm for extracting 
the field theory content and a parametrization of R- or global charges can be proposed \cite{aZequiv}:
\begin{itemize}
\item{Associate with every vertex $V_i$ of the toric diagram $P$ a real (positive) parameter $a_i$, $i=1,\ldots d$. Indexes $i$, $j$ are defined modulo $d$.
The parameters $a_i$ can be also displaced between the vectors of adjacent sides in the $(p,q)$ web, see Figure \ref{pqweb}.}
\item{Consider the set $C$ of all pairs of vectors $(v_i,v_j)$ in the $(p,q)$ web ordered in such a way  that the first vector $v_i$ can be rotated counter-clockwise to $v_j$ with an angle $\leq 180^o$.
Associate with every element of $C$ a type of chiral field in the field theory with multiplicity $|\langle v_i, v_j \rangle|$ and R-charge equal to $a_{i+1}+a_{i+2}+ \ldots a_{j}$. 
For example in Figure \ref{pqweb} the field associated to the pair $(v_d,v_3)$ has R-charge $a_1+a_2+a_3$ and multiplicity $|\langle v_d, v_3 \rangle|$. 

Note that eq. (\ref{numbers}) for $E$ is correctly reproduced.}
\item{Impose the linear constraint $\sum_{i=1}^d a_i = 2$
if you want to parametrize trial R-charges or
$\sum_{i=1}^d a_i =0$
if you want to parametrize the $d-1$ global charges.}
\end{itemize}

The chiral fields $\phi_i$ associated with the pair of consecutive vectors $(v_{i-1},v_i)$ with charge $a_i$ were already identified in \cite{tomorrow}: the dibaryon built with $\phi_i$ is dual to a D3 brane wrapped on the cycle $\Sigma_i$ associated with the vertex $V_i$.
The other chiral fields with charges obtained by summing consecutive $a_i$ form dibaryons dual to a D3 brane wrapped over the union of the corresponding $\Sigma_i$. The results in \cite{tomorrow} lead to a characterization of the
global charges: the $d-3$ baryonic charges are those that also satisfy $\sum_{i=1}^d a_i\, V_i=0$,
where $V_i$ are the coordinates of vertices in the plane of $P$.

With this assignment, we are able to write an expression for the trial $a$ function depending only on the toric diagram:
\begin{equation}
a=\frac{9}{32} \, \mathrm{tr}\, R^3 =\frac{9}{32} \left( V+ \sum_{(i,j) \in C}|\langle v_i, v_j \rangle| \,\, (a_{i+1}+a_{i+2}+\ldots a_{j}-1)^3 \right)
\label{aext}
\end{equation}   
Recall that $V$, the number of gauginos, is the double area of the polygon $P$ (\ref{numbers}).
In non minimal toric phases, which have additional fields, equation (\ref{aext}) is still true,
since the contribution from additional fields to the trial R-charge cancels \cite{aZequiv}. 

We can make several checks of this proposal for the field theory content. First of all, it is easy to
see that the proposal works in all cases where the quiver gauge theory is
explicitly known ($L^{p,q,r}$, $Y^{p,q}$, $X^{p,q}$, toric delPezzo). We also 
explicitly checked on many examples 
that the proposal is consistent with the general prescription given in \cite{dimers,rhombi} for extracting the quiver
gauge theory from the toric data (see the discussion in Section \ref{zigzag}).

We can also discuss the consistency of our proposal with the general
known properties of the $U(1)$ symmetries in this kind of theories.  
The proposal
reproduces, for example, the expected result ${\rm Tr} \, G=0$, where $G$ is a general R-charge or global symmetry charge. 
Moreover, it can be shown \cite{aZequiv}
that for the proposed assignment of charges the mixed cubic t'Hooft anomaly for baryonic symmetries is zero: ${\rm Tr}\, B_a^3=0$, as we expect from standard arguments in the gravitational dual \cite{intriwecht}.
  
However the best check of the proposal is that it allows to prove the equivalence of a-maximization and Z-minimization, as we shall soon discuss.

\section{Z-minimization}
\label{zeta}

It is in general hard to compute the Calabi-Yau metric corresponding to a particular toric diagram.
However in \cite{MSY} it was shown that all
the volumes we need can be computed from the toric data,
through the process known as volume minimization (or Z-minimization),
without any explicit knowledge of the metric. The reason for that is the following: supersymmetric cycles are calibrated and the volumes can be extracted
only from the Kahler form on the cone. The function Z to be minimized is a rational function of two variables $(x,y)$, which define a point inside the toric diagram.

Here we review the work of \cite{MSY}, reducing their formulas to the plane containing the convex polygon $P$.
Consider a generic toric diagram $P$ and a point $B=(x,y)$ allowed to vary inside $P$ \footnote{B is, up to a factor, the end point of the Reeb vector $K$, the vector field generating the U(1) isometry associated with the R-symmetry in field theory: $K = \sum _{i=1} ^{3} b_i e_i$, where $e_i$ form a basis for the $T^3$ fibration and $b=3(1,x,y)$.}. Call $r_i$ the vector joining $B$ with each vertex $V_i$.

Recall that every vertex $V_i$ of the toric diagram is associated with a calibrated submanifold $\Sigma_i$. Let us define the functions:
\begin{eqnarray}\label{volumes}
{\rm Vol}_{\Sigma_i}(x,y)&=&\frac{2\pi^2}{9} \frac{\langle v_{i-1},v_i\rangle}{\langle r_{i-1},v_{i-1}\rangle \langle
 r_{i},v_{i}\rangle}\equiv \frac{2\pi^2}{9} l_i(x,y)\nonumber \\
{\rm Vol}_H (x,y)&=&\frac{\pi}{6} \sum_{i=1}^d {\rm Vol}_{\Sigma_i}(x,y)
\end{eqnarray}
The function to minimize is just ${\rm Vol}_H (x,y)$. This is the function $Z$ in \cite{MSY} up to a constant multiplicative factor.
This function is convex inside $P$ and therefore has a unique minimum $(\bar x, \bar y)$; the volumes $Vol(\Sigma_i)$, $Vol(H)$ for the Calabi-Yau metric are given respectively by the values (\ref{volumes})  at the minimum 
\cite{MSY}.

\section{a-maximization is Z-minimization}
\label{comparison}

Now the problems of computing R-charges in field theory and volumes in the geometry
have been reduced to two different extremization problems defined only from toric data.
It is therefore possible to prove in the general toric case that the quantities they compute match according to the 
AdS/CFT predictions (\ref{central}) and (\ref{baryons}).
The general proof is given in \cite{aZequiv}; here we give only the main ideas and make useful observations.

To facilitate the comparison we define the geometrical function:
\begin{equation}
a^{MSY}(x,y)=\frac{\pi^3}{4 {\rm Vol}_H(x,y)}
\label{aMSY}
\end{equation}
and the functions:
\begin{equation}
f_i(x,y)= \frac{2 l_i(x,y)}{\sum_{j=1}^d l_j(x,y)}
\label{Rc}
\end{equation}
corresponding to the R-charges $R_i$ through equation (\ref{baryons}).
The process of Z-mi\-ni\-mi\-za\-tion can be restated as a maximization of $a^{MSY}(x,y)$ with respect to $(x,y)$ varying in the interior of $P$.

If we call $\bar a_i$ the values of $a_i$ at the local maximum, we have to prove that:
\begin{equation}
\begin{array}{l}
a^{MSY}(\bar x, \bar y) = a(\bar a_1, \bar a_2, \dots \bar a_d)\\[0.5em]
f_i(\bar x, \bar y)= \bar a_i \qquad i=1,\ldots d
\label{target}
\end{array}
\end{equation}
This is a highly non trivial check to perform:
a-maximization and Z-minimization use different functions and different trial charges;
it is not at all obvious why the result should be the same. 
First of all a-maximization is done on a total of $d-1$ independent
trial parameters while the volume minimization is done only on two parameters $(x,y)$. 
The trial central charge $a$ is a cubic polynomial in $a_i$, whereas 
$a^{MSY}$ is a rational function of $(x,y)$.
These parameters, in both cases, are somehow related to the possible global
symmetries: the Reeb vector in the geometry is connected to R-symmetries of the gauge theory 
and changing the position of $B$ in the directions $x$ and $y$ means adding to the R-symmetry
the two flavor global symmetries \footnote{Recall that flavor symmetries are mixed with baryonic ones,
so actually we are moving also in the space of baryonic symmetries.}. In any case, the volume minimization is done by moving only in a two dimensional
subspace of the set of global symmetries, while a-maximization is done
on the entire space: recall that the trial R-symmetry is mixed with both the two flavor symmetries and the $d-3$ baryonic ones. 

Therefore we try to decouple the baryon charges from the a-maximization
algorithm, performing a-maximization over a two dimensional subspace of parameters.
This subspace is just the space of coordinates $(x,y)$ of the plane where $P$ lies: consider the map from $\mathbb R^2$ to $\mathbb R^d$ given by
\begin{equation}
\begin{array}{l}
f:(x,y) \rightarrow (a_1,a_2,\ldots a_d)\\[1em]
\hspace{1.4em}(x,y) \rightarrow a_i= \displaystyle \frac{2 l_i(x,y)}{\sum_{j=1}^d l_j(x,y)}=f_i(x,y)
\label{map}
\end{array}
\end{equation} 
We are parameterizing the trial charges $a_i$ in field theory with the functions
(\ref{Rc}) taken at generic points $(x,y)$.

It is not difficult to prove that the gradient of the trial central charge along the $d-3$ baryonic directions evaluated on $f(P)$ is always zero:
\begin{equation}
\sum_{i=1}^d B^a_i \frac{\partial a}{\partial a_i}_{|a_i=f_i(x,y)}=0
\label{derbaryon}
\end{equation} 
where $B^a$ is a baryon charge and where the equality holds for every $(x,y)$ in the interior of $P$.
Therefore we have clarified in which sense the baryonic symmetries decouple from the process of a-maximization.

At this point we have to compare two functions of $(x,y)$: $a^{MSY}(x,y)$ and the field theory trial central charge evaluated 
on the surface $f(P)$. Remarkably one discovers that they are equal even before
maximization:
\begin{equation}
a(a_1,\ldots a_d)_{|a_i=f_i(x,y)}= a^{MSY}(x,y)
\label{equal}
\end{equation}
for every $(x,y)$ inside the interior of $P$. This shows the equivalence between a-maximization and Z-minimization.


\section{Distribution of charges in the dimer}
\label{zigzag}

The algorithm discussed in Section \ref{geometry} for extracting the gauge theory content from the toric diagram of the dual theory gives us the multiplicities and the charges of chiral fields, and this is enough to write equation (\ref{aext}) for the trial $a$ charge from the toric diagram, but it does not tell us how the chiral fields are disposed in the periodic quiver (or in the dimer). Recently a general prescription 
to extract the whole dimer from the toric diagram has been proposed 
\cite{rhombi}, \cite{mirror}; as we will soon explain this algorithm is based on zig-zag paths. Once the dimer is known, one can try to determine the distribution of trial R-charges and check the algorithm reviewed in Section \ref{geometry}. There are two equivalent ways of determining the distribution of R-charges, one already described in \cite{aZequiv} and a second one based on zig-zag paths. Here we will briefly discuss both of them and
show their equivalence. One can check on concrete examples that both prescription give the right distribution of R-charges: the sum of charges for a vertex in the dimer (superpotential term) is equal to 2, and the sum of charges of a face is equal to the number of sides minus 2 (beta function zero). 

The first general efficient way to find this distribution, valid for all toric phases, was discussed in \cite{aZequiv}: the parameters $a_i$ are associated with vertices of the toric diagram, and to every vertex $V_i$ there corresponds a single perfect matching\footnote{Recall that a perfect matching is a subset of links of the dimer configuration such that every white and black vertex is taken exactly once. Perfect matchings can be mapped to integer points of the toric diagram through the Kasteleyn matrix which counts their (oriented) intersections with two generators of the fundamental group of the torus \cite{dimers}.} in the dimer, at least for physical theories \cite{aZequiv,rhombi}. Therefore the trial charge of a link in the dimer can be computed as the sum of the parameters $a_i$ of all the external perfect matchings (corresponding to vertices) to which the link belongs. For examples
of how to use this prescription using the Kasteleyn matrix, see \cite{aZequiv}. 

We now propose another equivalent algorithm based on the results of \cite{rhombi} and \cite{mirror}: this will give new insight for the formula giving the multiplicities of fields.
\begin{figure}
\centering
\includegraphics[scale=0.65]{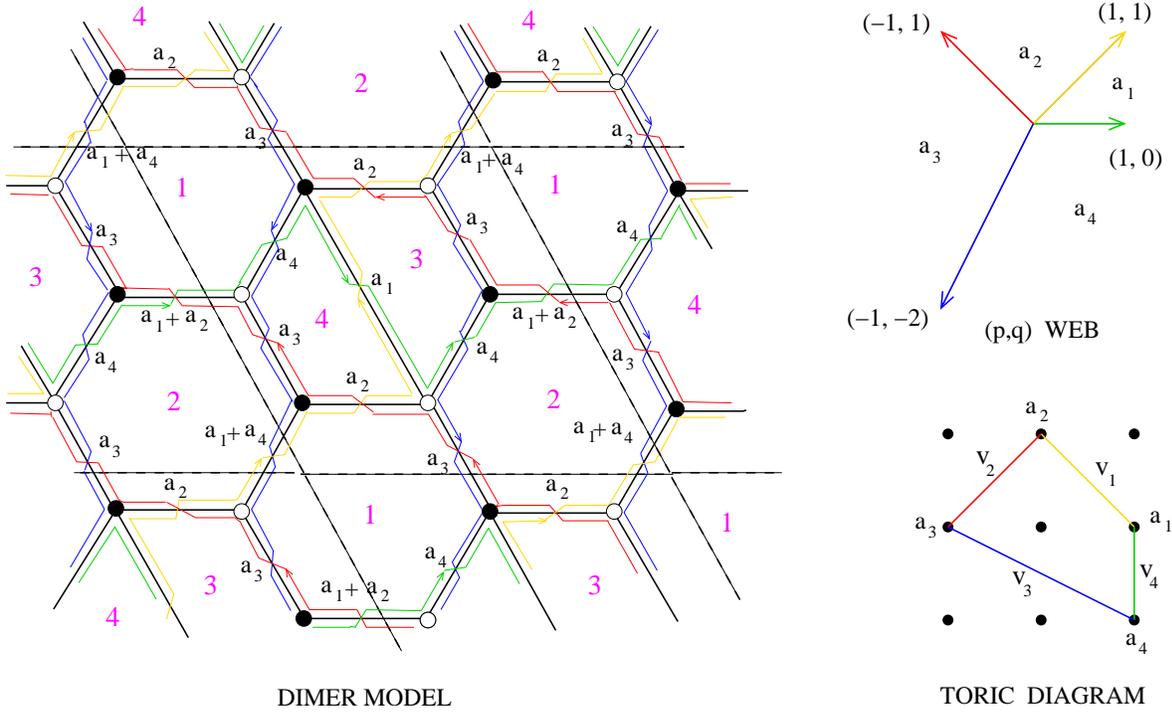}
\caption{The correspondence for $Y^{2,1}$.}
\label{y21}
\end{figure} 
A zig-zag path in the dimer is a path of links that turn maximally left at a node, maximally right at the next node, then again maximally left and so on \cite{rhombi}. We draw them in the specific case of $Y^{2,1}$ theory in Figure \ref{y21}: they are the four loops in green, yellow, red and blue and they are drawn so that they intersect in the middle of a link as in \cite{mirror}. Note that every link of the dimer belongs to exactly two different zig-zag paths, oriented in opposite directions. Moreover for dimer representing consistent theories the zig-zag paths are closed non-intersecting loops. There is a one to one correspondence between zig-zag path and legs of the $(p,q)$ web: the homotopy class in the fundamental group of the torus of every zig-zag path is just given by the integer numbers $(p,q)$ of the corresponding leg in the $(p,q)$ web \cite{rhombi}. The reader can check this directly in the example of Figure \ref{y21}: we can take as a fundamental cell for the torus one of the regions delimited by the black dashed lines.

Indeed the inverse algorithm of \cite{rhombi} consists just in drawing the zig-zag paths on a fundamental cell with the appropriate homotopy numbers. Since there's a one to one correspondence between intersections of zig-zag paths and links in the dimer, equation (\ref{numbers}) for the total number of fields is easily explained: recall that the number of topological intersections of two loops with homotopy class $w_i=(p_i,q_i)$ and $w_j=(p_j,q_j)$ is given by $\mathrm{det}(w_i,w_j)$. Since equation (\ref{numbers}) is valid in minimal toric phases of the gauge theory, we suggest that minimal phases can be built by making the effective number of intersections between every pair of closed zig-zag loops equal to the topological number. Non minimal phases, that is those with a greater number of fields than in (\ref{numbers}), may be explained with a greater number of effective intersection of zig-zag paths: we checked this in various examples. In the following we will concentrate on minimal toric phases.
   
The distribution of charges can be found as following. Consider the two zig-zag paths to which a link in the dimer belongs. They correspond to two vectors
$v_i=(p_i, q_i)$ and $v_j=(p_j, q_j)$ in the $(p,q)$ web. Then we propose that 
the charge of the link is given by the sum of the parameters $a_{i+1}+a_{i+2} \ldots +a_{j}$ between the vectors $v_i$ and $v_j$.  
So for instance in Figure \ref{y21} the links corresponding to the intersection of the red and the green zig-zag paths (vectors $v_4$ and $v_2$ in the (p,q) web) have charge equal to $a_1+a_2$. 

This rule explains the formula in Section \ref{geometry} for the multiplicities of fields with a given charge $a_{i+1}+a_{i+2} \ldots +a_{j}$: it counts the number of intersections between the zig zag paths corresponding to $v_i$ and $v_j$, which is just $\mathrm{det}(v_i,v_j)$.

To conclude, we explain the relation with the algorithm for charge distribution based on perfect matchings. We use a conjecture in \cite{rhombi}: if we consider the union of the two perfect matchings associated with the consecutive vertices $V_{j-1}$ and $V_{j}$  of $P$ we obtain the zig-zag path corresponding to the side $v_j$ plus some other isolated segments belonging to both the perfect matchings \footnote{note that we are also assuming that besides this segments and the zig-zag path there are no other closed loops with trivial homotopy.}: 
the zig-zag path is the ``formal'' difference of the two perfect matchings.
 Therefore if a link in the dimer belongs to the perfect matching $V_{j-1}$ but not to the perfect matching $V_{j}$ it belongs to the zig-zag path $v_j$; while if it belongs to both perfect matchings the link is not on the zig zag path $v_j$.
\begin{figure}
\centering
\includegraphics[scale=0.65]{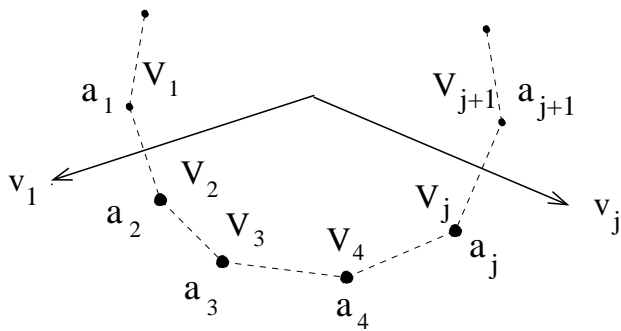}
\caption{Equivalence of the two algorithm for charge distribution.}
\label{proof}
\end{figure} 
With this assumption it is easy to prove the equivalence of the two algorithms. Consider a link in the dimer at the intersection of two zig-zag paths corresponding to the sides $v_1$ and $v_j$ of the $(p,q)$ web, as in Figure \ref{proof}. Since this link belongs to the zig-zag path $v_1$, it belongs to one of the perfect matchings $V_1$ or $V_2$ but not to both. Suppose it belongs to the perfect matching $V_2$. Then it is easy to see that it must belong also to the perfect matching $V_3$: otherwise the link would be in the zig zag path $v_2$, but this is not possible since the link must belong to exactly two zig-zag paths and it already belongs to the zig-zag paths $v_1$ and $v_j$. Continuing in this way one can prove that the link belongs to the perfect matchings (associated with) $V_2$, $V_3$, \ldots $V_j$, and so it will be given the charge $a_2+a_3\ldots +a_j$ with both algorithms. 
On the contrary if the link belongs to the perfect matching $V_1$ and not $V_2$, one derives that it belongs to all perfect matchings $V_1,V_d,\ldots V_{j+1}$: such fields may appear in non minimal toric phases \cite{aZequiv}.
This construction is also in agreement with the fact that chiral fields have always charges obtained by summing consecutive parameters $a_i$.

As an aside we note in Figure \ref{y21} that, taking into account also sums of $a_i$, for every vertex and face in the dimer the charges are in the cyclic order $a_1$, $a_2$, $a_3$, $a_4$ as in the toric diagram; vertices are all oriented anticlockwise, and faces are oriented clockwise. This seems to be a general fact.  

\vspace{3em}

\noindent {\Large{\bf Acknowledgments}}

\vspace{0.5em}

We thank Amihay Hanany, Sergio Benvenuti and Davide Forcella 
for useful discussions.
This work is supported in part by by INFN and MURST under 
contract 2001-025492, and by 
the European Commission TMR program HPRN-CT-2000-00131.

\end{document}